\documentclass{ws-ijgmmp}
\usepackage{colordvi}
\usepackage{color}

\begin{document}

\markboth{S. Capozziello,  M. De Laurentis}
{Noether Symmetries and Extended Gravity}

%
\catchline{}{}{}{}{}
%

\title{NOETHER SYMMETRIES IN EXTENDED GRAVITY QUANTUM COSMOLOGY}

\author{SALVATORE  CAPOZZIELLO and MARIAFELICIA DE LAURENTIS}

\address{Dipartimento di Fisica, Universita'
di Napoli {}``Federico II'' and
INFN Sezione  di Napoli, \\Compl. Univ. di
Monte S. Angelo, Edificio G, Via Cinthia, I-80126, Napoli, Italy.\\
E-mail: capozziello@na.infn.it; felicia@na.infn.it\\
www.unina.it}

\maketitle

\begin{history}
\received{(Day Month Year)}
\revised{(Day Month Year)}
\end{history}

\begin{abstract}
We summarize  the use of Noether symmetries in Minisuperspace Quantum Cosmology.  In particular, we consider  minisuperspace models, showing that the existence of conserved quantities   gives selection rules that allow to recover classical behaviors in cosmic evolution according to the so called Hartle criterion.  Such a criterion selects correlated regions in the configuration space of dynamical variables whose meaning is related to the emergence of classical observable universes. 
Some minisuperspace models are worked out starting from Extended  Gravity, in particular coming from  scalar tensor, $f(R)$ and $f(T)$ theories. 
Exact cosmological solutions are derived.\\

\noindent
 {\it We wish to dedicate this paper in loving memory of Mauro Francaviglia. We mourn  the loss of a  dear friend,  a peerless mentor and a traveling fellow in many scientific adventures, discussions and results.}

\end{abstract}

\keywords{Quantum Cosmology; Noether Symmetries; Minisuperspace Models.}

\section{Introduction}

The concept of symmetry has undergone  evolution and improvements over the last century: the transition from global continuous symmetries (such as translation and rotation) to  local continuous symmetries, and in particular local gauge symmetries, the addition of the internal symmetries to the space-time symmetries, the connection between charge conjugation and time reversal with the discrete symmetry of parity, the utilization of permutation symmetry in Quantum Field Theory and, finally, the use of symmetries in cosmology \cite{Branding,nuovocim}.
Symmetries have become a key tool of theory construction in fundamental physics. For example we recall the role of the Principle of Relativity in Einstein's construction of Special Relativity and then the role of general covariance as a symmetry principle in the development of General Relativity as well also the role of group theory in Weyl's approach to Quantum Field Theory or the development of particle physics and, for example, the use of symmetry principles in the construction of the Standard Model,  where symmetries play their role is the framework of Lagrangian field
theories, which are the current basis for any approach to fundamental interactions in Physics. This
corresponds, to consider a dynamical system with a continuous infinity of degrees of
freedom.

In this context, the work of Emmy Noether is of central importance since she provided a general proof that from the continuous symmetries of a given theory, we can derive the conservation laws of that theory.
In particular she discovered the relation between symmetries of dynamical  systems and their first integrals, {\it  i.e.}, physical
quantities which remain constant during the evolution of the system and are  related to fundamental
physical quantities such as energy, linear momentum, angular momentum and so on \cite{Noether,lorenzo}. 

Also in Quantum Cosmology the Noether symmetries have been of crucial importance, because they provide a subset of the general solution of the Wheeler-De Witt  equation where oscillating behaviors with crucial physical meaning can be  selected \cite{reviewhamilton}. Specifically, the Hartle criterion can be related to the Noether symmetries of the theory. In fact, in the context of the Minisuperspace Approach, this criterion  selects classical   trajectories which are   solutions of the  cosmological field equations \cite{halliwell}. In other words, the presence of Noether Symmetries in Quantum Cosmology allows the emergence of classically observable universes.

In this lecture notes, we will briefly sketch the Minisuperscape Approach in Quantum Cosmology (Sec. 2). Noether symmetries in Lagrangian and Hamiltonian  dynamical systems are discussed in Sec. 3. Applications to scalar-tensor, $f(R)$ and $f(T)$-gravity are extensively reported in Sec. 4. In Sec. 5 we draw  conclusions on the presented results.

\section{The Minisuperspace Approach to Quantum Cosmology}
\label{tre}

{\it Minisuperspaces}  are restrictions of superspace,  in which the symmetries are fixed a priori on the metric and matter fields.
The simplest model of minisuperspace is constructed by choosing
homogeneous and isotropic metrics and matter fields.
In general, it  provides a lapse function  $N=N(t)$ assumed homogeneous, the
shift functions  $N^i=0$ set to zero and a $4$-metric as follow
\begin{equation}
ds^2 = -N^2(t) dt^2 + h_{ij}({\bf x},t) dx^i dx^j\,. \label{5.1}
\end{equation}
  where  $h_{ij} $ is a $3$-metric
homogeneous and  it is described by a finite number of
functions of $t$, $q^{\alpha}(t)$, where $\alpha=0,1,2 \cdots (n-1)$(see \cite{reviewhamilton} and references therein).
We can recast the Hilbert-Einstein action as 
\begin{eqnarray}
{\cal S}[h_{ij},N,N^i] &=& {m_P^2 \over 16 \pi} \int dt \ d^3x \ N
\sqrt{h} \left[ K_{ij} K^{ij} -K^2+ {^{(3)} R} - 2 \Lambda \right]\,,
\label{5.5}
\end{eqnarray}
and, in general, one gets\footnote{Here,  we have  assumed as signature $ (-,+,+,+) $. The function  $ f_{\alpha\beta}(q) $ is the reduced  De Witt metric,
and the  integration of $t$  can range from $0$ to $1$ by
shifting $t$ and  rescaling the lapse function.}
 
\begin{eqnarray}
{\cal S}[q^{\alpha}(t), N(t)] &=& \int_0^1 dt N \left[ {1 \over 2N^2}
f_{\alpha\beta}(q) \dot q^{\alpha} \dot q^{\beta} - U(q) \right]\equiv \int {\cal L}
dt \label{5.6}\,.
\end{eqnarray}
the last equation has the form of  a relativistic point
particle action where the particles moves on a  $n$-dimensional  curved space-time  with a self-interaction
potential. The variation  with respect to $q^{\alpha}$, gives   the 
equations of motion
\begin{eqnarray}
{1 \over N} {d \over dt} \left( {\dot q^{\alpha} \over N} \right) + {1
\over N^2} \Gamma^{\alpha}_{\beta\gamma} \dot q^{\beta} \dot q^{\gamma} +
f^{\alpha\beta} {\partial U \over \partial q^{\beta} } = 0\,, \label{5.7}
\end{eqnarray}
where $\Gamma^{\alpha}_{\beta\gamma} $ are the  Christoffel symbols 
derived from the metric $f_{\alpha\beta}$. Varying with
respect to $N$, one gets 
\begin{eqnarray}
{1 \over 2 N^2} f_{\alpha\beta} \dot q^{\alpha} \dot q^{\beta} + U(q) = 0\,,
\label{5.8}
\end{eqnarray}
that is a constraint equation.

Eqs(\ref{5.7}) and (\ref{5.8}) describe geodesic motion in minisuperspace with a
forcing term.
The general solution of (\ref{5.7}), (\ref{5.8})
 requires $(2n-1)$ arbitrary parameters to be found.
In order to find the Hamiltonian,  the
canonical momenta have to be defined, that is 
\begin{eqnarray}
p_{\alpha} = {\partial {\cal L} \over \partial \dot q^{\alpha} } = f_{\alpha\beta} {\dot
q^{\beta} \over N}\,, \label{5.10}
\end{eqnarray}
and the canonical Hamiltonian is
\begin{eqnarray}
{\cal H}_c = p_{\alpha} \dot q^{\alpha} - {\cal L} = N \left[ \frac{1}{2} f^{\alpha\beta} p_{\alpha}
p_{\beta} + U(q) \right] \equiv N {\cal H}\,, \label{5.11}
\end{eqnarray}
where $f^{\alpha\beta}(q)$ is the inverse metric on minisuperspace. The action in 
Hamiltonian form is
\begin{eqnarray}
{\cal S} = \int_0^1 dt \left[ p_{\alpha} \dot q^{\alpha} - N{\cal H} \right]\,. \label{5.12}
\end{eqnarray}
A Lagrange
multiplier is  the lapse function $N$ and then the Hamiltonian constraint has to be
\begin{eqnarray}
{\cal H}(q^{\alpha},p_{\alpha}) = \frac{1}{2} f^{\alpha\beta} p_{\alpha} p_{\beta} + U(q) = 0\,.
\label{5.13}
\end{eqnarray}

Now, the
canonical quantization  procedure requires  a
time-independent wave function $ \Psi(q^{\alpha}) $  that has to be
 annihilated by the quantum operator corresponding to the classical
constraint (\ref{5.13}). This fact gives rises to the the Wheeler-De Witt   equation,
\begin{eqnarray}
\hat {\cal H}(q^{\alpha}, -i {\partial \over \partial q^{\alpha} } ) \Psi(q^{\alpha})
= 0\,, \label{5.14}
\end{eqnarray}
where $\Psi(q^{\alpha})$ is the so-called {\it Wave Function of the Universe}. 
Since the metric $f^{\alpha\beta}$ depends on $q$ there is a
factor ordering issue in (\ref{5.14}). This may be
solved by requiring that the quantization procedure is
covariant in minisuperspace,  that is unchanged by
field redefinitions of the $3$-metric and matter fields, $q^{\alpha}
\to \tilde q^{\alpha}(q^{\alpha})$. This fact restricts the possible
operator orderings to
\begin{eqnarray}
\hat {\cal H} = - \frac{1}{2} \nabla^2 + \xi {\cal R} + U(q)\,, \label{5.15}
\end{eqnarray}
where $ \nabla^2 $ and $ {\cal R} $ are the Laplacian and curvature of
the minisuperspace metric $f_{\alpha\beta}$ and $\xi$ is an arbitrary
constant.

%

Before concluding this section, an important issue has to be addressed. It is how to interpret the probability measure  in Quantum Cosmology.
As a final remarks we remember as we can interpret the probability measure. 
Since the  Wheeler-De Witt  equation is similar to Klein-Gordon  equation we can define a current,  that is conserved and satisfies $\nabla \cdot J = 0$, in this way 
\begin{eqnarray}
J = {i \over 2 } \left( \Psi^* \nabla \Psi - \Psi \nabla \Psi^*
\right)\,. \label{5.26}
\end{eqnarray}
 As in the case  of the Klein-Gordon
equation (and, in general, of hyperbolic equations), the probability derived from such a 
conserved current can be affected by   negative
probabilities. Due to this shortcoming, 
the correct measure to use should be
\begin{eqnarray}
dP = |\Psi(q^{\alpha})|^2 dV\,, \label{5.28}
\end{eqnarray}
where $dV$ is a volume element of minisuperspace \cite{reviewhamilton,halliwell}. 
%

\section{The Noether Symmetry Approach}
\label{quattro}
As we said before, minisuperspaces are restrictions of the superspace of
geometrodynamics. They are finite-dimensional configuration spaces
on which point-like Lagrangians can be defined.
Cosmological models of physical interest can be defined on such
minisuperspaces ({\it e.g.} Bianchi models). According to the above discussion, 
a crucial role is played by the conserved currents that allow to interpret the probability 
measure and then the physical quantities obtained in Quantum Cosmology. In this context, the search for general methods to achieve conserved quantities and symmetries become  relevant. The so-called {\it Noether Symmetry Approach} \cite{nuovocim}, as we will show, can be extremely useful to this purpose.

Let ${\cal
L}$ a Lagrangian defined on the tangent space of
configurations $T{\cal Q}\equiv\{q_i, \dot{q}_i\}$, 
the vector
field $X$ is 
 \begin{equation}\label{17}
 X=\alpha^i(q)\frac{\partial}{\partial q^i}+
 \dot{\alpha}^i(q)\frac{\partial}{\partial\dot{q}^i}\,{,}
 \end{equation}
 where dot means derivative with respect to $t$, and
 \begin{equation}\label{18}
 L_X{\cal L}=X{\cal L}=\alpha^i(q)\frac{\partial {\cal L}}{\partial q^i}+
 \dot{\alpha}^i(q)\frac{\partial {\cal L}}{\partial\dot{q}^i}\,{,}
 \end{equation}
 where is the Lie derivative $L_X$ of ${\cal L}$.
 The condition
 \begin{equation}\label{19}
 L_X{\cal L}=0\,,
 \end{equation}
 implies that the phase flux is conserved along $X$: this means that a
constant of motion exists for ${\cal L}$ and the Noether theorem
holds. In fact, taking into account the Euler-Lagrange equations
 \begin{equation}\label{20}
 \frac{d}{dt}\frac{\partial {\cal L}}{\partial\dot{q}^i}-
 \frac{\partial {\cal L}}{\partial q^i}=0\,{,}
 \end{equation}
 it is easy to show that
 \begin{equation}\label{21}
 \frac{d}{dt}\left(\alpha^i\frac{\partial {\cal
 L}}{\partial\dot{q}^i}\right)=L_X{\cal L}\,{.}
 \end{equation}
 If Eq.(\ref{19}) holds,
 \begin{equation}\label{22}
 \Sigma_0=\alpha^i\frac{\partial {\cal L}}{\partial\dot{q}^i}
 \end{equation}
 is a constant of motion. Alternatively, using the Cartan
one--form
 \begin{equation}\label{23}
 \theta_{{\cal L}}\equiv\frac{\partial {\cal
 L}}{\partial\dot{q}^i}\,dq^i
 \end{equation}
 and defining the inner derivative
 \begin{equation}\label{24}
 i_X\theta_{{\cal L}}=<\theta_{{\cal L}}, X>\,{,}
 \end{equation}
 we get, as above,
 \begin{equation}\label{25}
 i_X\theta_{{\cal L}}=\Sigma_0\,,
 \end{equation}
 if condition (\ref{19}) holds.
 This representation is useful to identify cyclic variables. Using
a point transformation on vector field (\ref{17}), it is possible
to get\footnote{We indicate the quantities as
Lagrangians and  vector fields with a tilde if the
non--degenerate transformation
 $$
 Q^i=Q^i(q)\,{,}\quad \dot{Q}^i(q)=\frac{\partial Q^i}{\partial
 q^j}\,\dot{q}^j\,,
 $$
 is performed. 
 However the Jacobian determinant
 ${\cal J}=\parallel\partial Q^i
 /\partial q^j\parallel$ has to be non--zero.}
 \begin{equation}\label{26}
 \tilde{X}=(i_XdQ^k)\,\frac{\partial}{\partial Q^k}+
 \left[\frac{d}{dt}(i_XdQ^k)\right]\frac{\partial}{\partial\dot{Q}^k}\,{.}
 \end{equation}
 If $X$ is a symmetry also $\tilde{X}$ has this property, then it
is always possible to choose a coordinate transformation so that
 \begin{equation}\label{27}
 i_XdQ^1=1\,{,} \quad i_XdQ^i=0\,{,}\quad i\neq 1\,{,}
 \end{equation}
 and then
 \begin{equation}\label{28}
 \tilde{X}=\frac{\partial}{\partial Q^1}\,{,}\quad
 \frac{\partial\tilde{{\cal L}}}{\partial Q^1}=0\,{.}
 \end{equation}
 It is evident that $Q^1$ is the cyclic coordinate and the
dynamics can be reduced \cite{arnold}. However, the change of
coordinates is not unique and a clever choice is always important.
 Furthermore, it is possible that more symmetries are found.
In this case,  more than one cyclic variable exists. 

%
A reduction procedure by cyclic coordinates can be implemented in
three steps:
\begin{itemize}
\item  we choose a symmetry and obtain new coordinates as
above. After this first reduction, we get a new Lagrangian
$\tilde{{\cal L}}$ with a cyclic coordinate; 
\item  we search for new
symmetries in this new space and apply the reduction technique
until it is possible; \item 
 the process stops if we select a pure
kinetic Lagrangian where all coordinates are cyclic.
\end{itemize}

 Going
back to the point of view interesting in Quantum Cosmology, any
symmetry selects a constant conjugate momentum since, by the
Euler-Lagrange equations
 \begin{equation}\label{32}
 \frac{\partial\tilde{{\cal L}}}{\partial Q^i}=0\Longleftrightarrow
 \frac{\partial\tilde{{\cal L}}}{\partial \dot{Q}^i}=\Sigma_i\,{.}
 \end{equation}
 Viceversa, the existence of a constant conjugate momentum means
that a cyclic variable has to exist. In other words, a Noether
symmetry exists.

Further remarks on the form of the Lagrangian ${\cal L}$ are
necessary at this point. We shall take into account
time--independent, non--degenerate Lagrangians ${\cal L}={\cal
L}(q^i, \dot{q}^j)$, {\it i.e.}
 \begin{equation}\label{33}
 \frac{\partial {\cal L}}{\partial t}=0\,{,}\quad
 \mbox{det} H_{ij}\equiv \mbox{det}\vert\vert\frac{\partial^2{\cal L}}
 {\partial\dot{q}^i\partial\dot{q}^j}\vert\vert \neq 0\,{,}
 \end{equation}
 where $H_{ij}$ is the Hessian.
 As in usual analytic mechanics, ${\cal L}$ can be set in the form
 \begin{equation}\label{34}
 {\cal L}=T(q^i, \dot{q}^i)-V(q^i)\,{,}
 \end{equation}
 where $T$ is a positive--defined quadratic form in the
$\dot{q}^j$ and $V(q^i)$ is a potential term. The energy function
associated with ${\cal L}$ is
 \begin{equation}\label{35}
 E_{{\cal L}}\equiv \frac{\partial {\cal L} }{\partial
 \dot{q}^i}\,\dot{q}^i-{\cal L}(q^j, \dot{q}^j)\,,
 \end{equation}
 and by the Legendre transformations
 \begin{equation}\label{36}
 {\cal H}=\pi_j\dot{q}^j-{\cal L}(q^j, \dot{q}^j)\,{,}\quad
 \pi_j=\frac{\partial {\cal L}}{\partial \dot{q}^j}\,{,}
 \end{equation}
 we get the Hamiltonian function and the conjugate momenta.
Considering again the symmetry, the condition (\ref{19}) and the
 vector field $X$ in Eq.(\ref{17}) give a homogeneous polynomial
of second degree in the velocities plus an inhomogeneous term in
the $q^j$. Due to (\ref{19}), such a polynomial has to be
identically zero and then each coefficient must be independently
zero. If $n$ is the dimension of the configuration space ({\it i.e.} the
dimension of the minisuperspace), we get $\{ 1+n(n+1)/2\}$ partial
differential equations whose solutions assign the symmetry, as we
shall see below. Such a symmetry is over--determined and, if a
solution exists, it is expressed in terms of integration constants
instead of boundary conditions.
In the Hamiltonian formalism, we have
 \begin{equation}\label{43}
 [\Sigma_j, {\cal H}]=0\,{,}\quad 1\leq j \leq m\,{,}
 \end{equation}
 as it must be for conserved momenta in quantum mechanics and the
Hamiltonian has to satisfy the relations
 \begin{equation}\label{44}
 L_{\Gamma}{\cal H}=0\,{,}
 \end{equation}
in order to obtain a Noether symmetry. The vector $\Gamma$ is defined by
\cite{marmo}
 \begin{equation}\label{45}
 \Gamma =\dot{q}^i\frac{\partial}{\partial q^i}+
 \ddot{q}^i\frac{\partial}{\partial\dot{q}^i}\,{.}
 \end{equation}
These considerations can be applied to the minisuperspace models of  Quantum Cosmology and to the
 interpretation of the wave function of the universe.

As discussed above, by a straightforward canonical quantization procedure, we have
 \begin{eqnarray}
 \pi_j & \longrightarrow & \hat{\pi}_j=-i\partial_j\,{,} \label{37} \\
 {\cal H} & \longrightarrow & \hat{\cal H}(q^j,
 -i\partial_{q^j})\,{.}\label{38a}
 \end{eqnarray}
It is well known that the Hamiltonian constraint gives the Wheeler-De Witt 
equation, so that if $\vert \Psi>$ is a {\it state} of the system
({\it i.e.} the wave function of the universe), dynamics is given by
 \begin{equation}\label{39a}
 {\cal H}\vert\Psi>=0\,{,}
 \end{equation}
 where we write the Wheeler-De Witt  equation in an operational  way.
 If a Noether symmetry exists, the reduction procedure outlined
above can be applied and then, from (\ref{32}) and (\ref{36}), we
get
 \begin{eqnarray}
 \pi_1\equiv\frac{\partial {\cal
 L}}{\partial\dot{Q}^1}=i_{X_1}\theta_{{\cal L}}& = &
 \Sigma_1\,{,}\nonumber \\
 \pi_2\equiv\frac{\partial {\cal
 L}}{\partial\dot{Q}^2}=i_{X_2}\theta_{{\cal L}}& = &
 \Sigma_2\,{,}\label{40a} \\
\ldots\quad \ldots & & \ldots \,{,} \nonumber
 \end{eqnarray}
 depending on the number of Noether symmetries. After
quantization, we get
 \begin{eqnarray}
 -i\partial_1\vert\Psi>&=&\Sigma_1\vert\Psi>\,{,} \nonumber \\
 -i\partial_2\vert\Psi>&=&\Sigma_2\vert\Psi>\,{,} \label{41a} \\
 \ldots & & \ldots \,{,} \nonumber
 \end{eqnarray}
which are nothing else but translations along the $Q^j$ axis
singled out by corresponding symmetry. Eqs. (\ref{41a}) can be
immediately integrated and, being $\Sigma_j$ real constants, we
obtain oscillatory behaviors for $\vert \Psi>$ in the directions
of symmetries, {\it i.e.}
 \begin{equation}\label{42a}
 \vert\Psi>=\sum_{j=1}^m\, e^{i\Sigma_jQ^j}\vert \chi(Q^l)>\,{,}
 \quad m < l\leq n\,{,}
 \end{equation}
 where $m$ is the number of symmetries, $l$ are the directions
where symmetries do not exist, $n$ is the total dimension of
minisuperspace.
Viceversa, dynamics given by (\ref{39a}) can be reduced by
(\ref{41a}) if and only if it is possible to define constant
conjugate momenta as in (\ref{40a}), that is oscillatory behaviors
of a subset of solutions $\vert\Psi>$ exist only if Noether
symmetry exists for dynamics.

The $m$ symmetries give first integrals of motion and then the
possibility to select classical trajectories. In one and
two--dimensional minisuperspaces, the existence of a Noether
symmetry allows the complete solution of the problem and to get
the full semi-classical limit of Quantum Cosmology \cite{minisup}.
In conclusion, we can state that
in the semi-classical limit of quantum
cosmology, the reduction procedure of dynamics, connected to the
existence of Noether symmetries, allows to select a subset of the
solution of Wheler De Witt equation where oscillatory behaviors are found.
This fact, in the framework of the Hartle interpretative
criterion of the wave
function of the universe, gives conserved momenta and trajectories
which can be interpreted as classical cosmological solutions.
Vice-versa, if a subset of the solution of Wheeler-De Witt  equation has an
oscillatory behavior, due to Eq.(\ref{19}), conserved momenta 
exist and Noether symmetries are present. In other words, {\it Noether
symmetries select classical universes and then are directly related to  the validity of the Hartle criterion}.

In what follows, we will show that such a statement holds for general classes of minisuperspaces and allows to select exact classical solutions, {\it i.e.} the presence of Noether symmetries is a selection criterion for classical universes. In  the next section, we will shown how they work for Extended Theories of Gravitaty, as scalar tensor,  $f(R)$ and $f(T)$ gravity.

\section{ Noether symmetries in cosmology}
Let us consider realizations of the above approach in  minisuperspace cosmological
models derived from Extended  Theories of Gravity. 
As we have seen, the Hartle criterion is directly connected to the presence of  Noether symmetries since oscillatory behaviors  means correlations among variables \cite{PROD,review1,review2,physrep,physrep1}.

Specifically, the approach can be  connected  to the search for Lagrange multipliers.
In fact, imposing  Lagrange
multipliers allow to  modify the
dynamics and select the form of  effective potentials.
By integrating the multipliers, 
solutions can be obtained.

On the other hand,  the Lagrange multipliers are constraints capable of
reducing  dynamics. Technically they
are anholonomic constraints being time-dependent. They give rise
to field equations which describe  dynamics of the further
degrees of freedom coming from Extended  Theories of Gravity \cite{PROD,review1,review2,physrep,physrep1}. This fact is
extremely relevant to deal with  new degrees of freedom under the
standard of effective scalar fields \cite{moltiplicatori}.
Below,   we give minisuperspace examples and obtain  exact cosmological solutions.
In particular, we show that, by imposing Lagrange multipliers, a given  minisuperspace model becomes canonical and Noether symmetries, if exist, can be found out.

\subsection{Scalar-Tensor  cosmology}
%
A general action for a nonminimally coupled theory of gravity
assume the following form
 \begin{equation}\label{46a}
 {\cal S}=\int d^4x\sqrt{-g}\left[F(\phi)R+\frac{1}{2}g^{\mu\nu}
 \phi_{\mu}\phi_{\nu}-V(\phi)\right]\,{,}
 \end{equation}
 where, $F(\phi)$ and $V(\phi)$ are respectively the coupling
and the potential of a scalar field\footnote{Using
physical units $8\pi G=c=\hbar =1$, the standard Einstein
coupling is recovered for $F(\phi)=-1/2$.}.

The cosmological point-like Lagrangian for a  Friedman Ð Robertson Ð Walker (FRW) minisuperspace  in terms of the scale factor $a$ is 
 \begin{equation}\label{47a}
 {\cal L}=6a\dot{a}^2F+6a^2\dot{a}\dot{F}-
 6kaF+a^3\left[\frac{\dot{\phi}}{2}-V\right]\,{.}
 \end{equation}

The configuration space of such a Lagrangian is ${\cal Q}\equiv\{a,
\phi\}$, {\it i.e.} a bidimensional minisuperspace. A Noether
symmetry exists if Eq. (\ref{19}) holds. In this case, it has to be
 \begin{equation}\label{48a}
 X=\alpha\, \frac{\partial}{\partial a}+
 \beta\,\frac{\partial}{\partial \phi}+
 \dot{\alpha}\,\frac{\partial}{\partial\dot{a}}+
 \dot{\beta}\,\frac{\partial}{\partial\dot{\phi}}\,{,}
 \end{equation}
where $\alpha, \beta$ depend on $a, \phi$. This vector field acts on the ${\cal Q}$ minisuperspace. The system of
partial differential equation given by (\ref{19}) is
 \begin{eqnarray}a
&& F(\phi)\left[\alpha+2a\frac{\partial\alpha}{\partial
 a}\right]+
 a F'(\phi)\left[\beta+a\frac{\partial\beta}{\partial
 a}\right] =  0\,{,} \label{49a}\\
&& 3\alpha+12F'(\phi)\frac{\partial\alpha}{\partial\phi}+2a
 \frac{\partial\beta}{\partial\phi}  =  0\,{,}\label{50a} \\
&& a\beta F''(\phi)+\left[2\alpha+a\frac{\partial \alpha}{\partial a}
 +\frac{\partial\beta}{\partial\phi}\right]F'(\phi)+2\frac{\partial\alpha}{\partial\phi}F(\phi)+\frac{a^2}{6}
 \frac{\partial\beta}{\partial a} =  0\,{,} \label{51a} \\
&& [3\alpha V(\phi)+a\beta V'(\phi)]a^2+6k[\alpha F(\phi)+
 a\beta F'(\phi)]  =  0\,{.}\label{52a}\nonumber\\
 \end{eqnarray}
Prime indicates the derivative with respect to $\phi$. The
number of equations is $4$ as it has to be, being $n=2$ the $\cal Q$-dimension. Several
solutions exist for this system \cite{pla,cqg,prd}. They determine
also the form of the model since the system (\ref{49a})-(\ref{52a})
gives $\alpha, \beta$, $F(\phi)$ and $V(\phi)$. For example,
if the spatial curvature is $k=0$, a solution is
 \begin{equation}\label{53a}
 \alpha=-\frac{2}{3}p(s)\beta_0a^{s+1}\phi^{m(s)-1}\,{,}\quad
 \beta=\beta_0a^s\phi^{m(s)}\,{,}
 \end{equation}
 \begin{equation}\label{54a}
 F(\phi)=D(s)\phi^2\,{,}\quad
 V(\phi)=\lambda\phi^{2p(s)}\,{,}
 \end{equation}
 where
 \begin{eqnarray}\label{55a}
 D(s)&=&\frac{(2s+3)^2}{48(s+1)(s+2)}\,{,}\nonumber\\
 \\
 p(s)&=&\frac{3(s+1)}{2s+3}\,{,}\nonumber\\
 \\
 m(s)&=&\frac{2s^2+6s+3}{2s+3}\,{,}\nonumber\\
 \end{eqnarray}
 and $s, \lambda$ are free parameters. The change of variables
(\ref{27}) gives
 \begin{equation}\label{56a}
 w=\sigma_0a^3\phi^{2p(s)}\,{,}\quad z=\frac{3}{\beta_0\chi(s)}
 a^{-s}\phi^{1-m(s)}\,{,}
 \end{equation}
 where $\sigma_0$ is an integration constant and
 \begin{equation}\label{57a}
 \chi(s)=-\frac{6s}{2s+3}\,{.}
 \end{equation}
 Lagrangian (\ref{47a}) becomes, for $k=0$,
 \begin{equation}\label{58a}
 {\cal L}=\gamma(s)w^{s/3}\dot{z}\dot{w}-\lambda w\,{,}
 \end{equation}
 where $z$ is cyclic and
 \begin{equation}\label{59a}
 \gamma(s)=\frac{2s+3}{12\sigma_0^2(s+2)(s+1)}\,{.}
 \end{equation}
 The conjugate momenta are
 \begin{equation}\label{60a}
 \pi_z=\frac{\partial {\cal L}}{\partial \dot{z}}=\gamma(s)w^{s/3}
 \dot{w}\,{,}\quad
 \pi_w=\frac{\partial {\cal L}}{\partial \dot{w}}=\gamma(s)w^{s/3}
 \dot{z}\,{,}
 \end{equation}
 and the Hamiltonian is
 \begin{equation}\label{61a}
 \tilde{{\cal
 H}}=\frac{\pi_z\pi_w}{\gamma(s)w^{s/3}}+\lambda w\,{.}
 \end{equation}
 The Noether symmetry is given by
 \begin{equation}\label{62a}
 \pi_z=\Sigma_0\,{.}
 \end{equation}
 Quantizing Eqs. (\ref{60a}), we  get
 \begin{equation}\label{63}
 \pi\longrightarrow -i\partial_z \,{,}\quad \pi_w\longrightarrow
 -i\partial_w\,{,}
 \end{equation}
 and then the Wheeler-De Witt  equation
 \begin{equation}\label{64a}
 [(i\partial_z)(i\partial_w)+\tilde{\lambda}w^{1+s/3}]\vert\Psi>=0\,{,}
 \end{equation}
 where $\tilde{\lambda}=\gamma(s)\lambda$.
The quantum version of constraint (\ref{62a}) is
 \begin{equation}\label{65a}
 -i\partial_z\vert\Psi>=\Sigma_0\vert\Psi>\,{,}
 \end{equation}
 so that dynamics results reduced. A straightforward integration
 of Eqs. (\ref{64a}) and (\ref{65a}) gives
  \begin{equation}\label{66a}
  \vert\Psi>=\vert\Omega(w)>\vert\chi(z)>\propto
  e^{i\Sigma_0z}\,e^{-i\tilde{\lambda}w^{2+s/3}}\,{,}
  \end{equation}
which is an oscillating wave function and the Hartle criterion is
recovered. In the semi--classical limit, we have two first integrals
of motion: $\Sigma_0$ ({\it i.e.} the equation for $\pi_z$) and
$E_{{\cal L}}=0$,{\it  i.e.} the Hamiltonian (\ref{61a}) which becomes
the equation for $\pi_{w}$. Classical trajectories
in
the configuration space $\tilde{\cal Q}\equiv\{w, z\}$ are immediately
recovered
 \begin{eqnarray}
 w(t)&=&[k_1t+k_2]^{3/(s+3)}\,{,}\label{67} \\
 z(t)&=&[k_1t+k_2]^{(s+6)/(s+3)}+z_0\,{,}\label{68a}
 \end{eqnarray}
then, going back to ${\cal Q}\equiv\{a, \phi\}$,
we get the {\it classical}
cosmological behaviour
 \begin{eqnarray}
 a(t)&=&a_0(t-t_0)^{l(s)}\,{,} \label{69} \\
 \phi(t)&=&\phi_0(t-t_0)^{q(s)}\,{,} \label{70a}
 \end{eqnarray}
 where
 \begin{equation}\label{71a}
 l(s)=\frac{2s^2+9s+6}{s(s+3)}\,{,}\quad q(s)=-\frac{2s+3}{s}\,{,}
 \end{equation}
 which means that Hartle criterion selects classical universes.
Depending on the value of $s$, we get Friedman, power--law, or
pole--like behaviors. 
The
considerations on the oscillatory regime of the wave function of
the universe and the recovering of classical behaviors are
exactly the same.

\subsection{$f(R)$ cosmology}
Similar results can be obtained also for higher--order gravity minisuperspaces. In
particular, let us consider fourth--order gravity given by the
action
 \begin{equation}\label{72a}
 {\cal S}=\int d^4x \sqrt{-g}\, f(R)\,{,}
 \end{equation}
where $f(R)$ is a generic function of Ricci scalar \cite{physrep}. 
 Rewriting the action to a
point-like FriedmanÐRobertsonÐWalker one, we obtain
 \begin{equation}\label{73a}
 {\cal S}=\int dt {\cal L}(a, \dot{a}; R, \dot{R})\,{,}
 \end{equation}
where dot means derivative with respect to the cosmic time. The
scale factor $a$ and the Ricci scalar $R$ are the canonical
variables. This position could seem arbitrary since $R$ depends on
$a, \dot{a}, \ddot{a}$, but it is generally used in canonical
quantization \cite{vilenkin1,Schmidt,lambda}. The
definition of $R$ in terms of $a, \dot{a}, \ddot{a}$ introduces a
constraint which eliminates second and higher order derivatives in
action (\ref{73a}), and yields to a system of second order
differential equations in $\{a, R\}$. Action (\ref{73a}) can be
written as
 \begin{equation}\label{74a}
 {\cal S}=2\pi^2\int dt \left\{ a^3f(R)-\lambda\left [ R+6\left (
 \frac{\ddot{a}}{a}+\frac{\dot{a}^2}{a^2}+\frac{k}{a^2}\right)\right]\right\}\,{,}
 \end{equation}
where the Lagrange multiplier $\lambda$ is derived by varying
with respect to $R$. It is
 \begin{equation}\label{75a}
 \lambda=a^3f'(R)\,{.}
 \end{equation}
 Here prime means derivative with respect to $R$. To recover the
 analogy with previous scalar--tensor models, let us
 introduce the auxiliary field
 \begin{equation}\label{76a}
  p\equiv f'(R)\,{,}
 \end{equation}
 so that the Lagrangian in (\ref{74a}) becomes
 \begin{equation}\label{77a}
 {\cal L}=6a\dot{a}^2p+6a^2\dot{a}\dot{p}-6kap-a^3W(p)\,{,}
 \end{equation}
 which is of the same form of (\ref{47a}) a part the kinetic term.
This is an Helmhotz--like Lagrangian \cite{magnano} and $a, p$ are
independent fields. The potential $W(p)$ is defined as
 \begin{equation}\label{78a}
 W(p)=h(p)p-r(p)\,{,}
 \end{equation}
where
 \begin{equation}\label{rh(p)}
 r(p)=\int f'(R)dR=\int pdR=f(R)\,{,} \quad h(p)=R\,{,}
 \end{equation}
such that $h=(f')^{-1}$ is the inverse function of $f'$. The
configuration space is now ${\cal Q}\equiv\{a, p\}$ and $p$ has
the same role
of the above $\phi$. Condition (\ref{19}) is now realized by
the vector field
 \begin{equation}\label{80a}
 X=\alpha (a, p)\frac{\partial}{\partial a}+\beta(a, p)
 \frac{\partial}{\partial p}+\dot{\alpha}\frac{\partial}{\partial\dot{a}}
 +\dot{\beta}\frac{\partial}{\partial\dot{p}}\,,
 \end{equation}
 and explicitly it gives the system
 \begin{eqnarray}a
&&p\left[\alpha+2a\displaystyle\frac{\partial\alpha}{\partial a}\right]p+
a\left[\beta+a\displaystyle\frac{\partial\beta}{\partial a}\right] =
 0\,{,}\label{81}  \\
&& a^2\displaystyle \frac{\partial\alpha}{\partial p}=0\,{,} \label{82} \\
&& 2\alpha+a\displaystyle \frac{\partial\alpha}{\partial
 a}+2p\displaystyle \frac{\partial\alpha}{\partial p}+
a \frac{\partial\beta}{\partial p}=0\,{,}\label{83} \\
&& 6k[\alpha p+\beta a]+a^2[3\alpha W+
a \beta \displaystyle\frac{\partial W}{\partial
 p}]=0\,{.}\label{84a}
\end{eqnarray}
 The solution of this system, {\it i.e.} the existence of a Noether
symmetry, gives $\alpha$, $\beta$ and $W(p)$. It is satisfied for
\begin{equation}\label{85a}
  \alpha=\alpha (a)\,{,} \qquad \beta (a, p)=\beta_0 a^sp\,{,}
 \end{equation}
where $s$ is a parameter and $\beta_0$ is an integration constant.
In particular,
 \begin{eqnarray}\label{86a}
 && s=0\longrightarrow \alpha (a)=-\frac{\beta_0}{3}\, a\,{,} \quad
 \beta (p)=\beta_0\, p\,{,} \nonumber\\
 \nonumber\\ && \quad W(p)=W_0\, p\,{,} \quad
 k=0\,{,}
 \end{eqnarray}
 \begin{eqnarray}\label{87a}
 && s=-2\longrightarrow \alpha (a)=-\frac{\beta_0}{a}\,{,}\quad
 \beta (a, p) = \beta_0\, \frac{p}{a^2}\,{,} \quad\nonumber\\
 \nonumber\\&&W(p)=W_1p^3\,{,} \quad \forall \,\,\, k\,{,}
 \end{eqnarray}
 where $W_0$ and $W_1$ are constants. 
Let us discuss separately the
solutions (\ref{86a}) and (\ref{87a}).

\subsubsection{The case $s=0$}

The induced change of variables
 $\displaystyle{{\cal Q}\equiv \{a, p\}
 \longrightarrow \tilde{Q}\equiv \{w, z\}}$
 can be
 \begin{equation}\label{88a}
 w(a, p)=a^3p\,{,} \quad z(p)=\ln p\,{.}
 \end{equation}
 Lagrangian (\ref{77a}) becomes
 \begin{equation}\label{89}
 \tilde{{\cal L}}(w, \dot{w},
 \dot{z})=\dot{z}\dot{w}-2w\dot{z}^2+\frac{\dot{w}^2}{w}-3W_0w\,{.}
 \end{equation}
 and, obviously, $z$ is the cyclic variable. The conjugate momenta are
 \begin{equation}\label{90a}
 \pi_z\equiv\frac{\partial \tilde{{\cal L}}}{\partial
 \dot{z}}=\dot{w}-4\dot{z}=\Sigma_0\,{,}
 \end{equation}
 \begin{equation}\label{91a}
 \pi_w\equiv\frac{\partial \tilde{{\cal L}}}{\partial
 \dot{w}}=\dot{z}+2\frac{\dot{w}}{w}\,{.}
 \end{equation}
 and the Hamiltonian is
 \begin{equation}\label{92a}
 {\cal H}(w, \pi_w, \pi_z)=
 \pi_w\pi_z-\frac{\pi_z^2}{w}+2w\pi^2_w+6W_0w\,{.}
 \end{equation}
 By canonical quantization, reduced dynamics is given by
 \begin{equation}\label{93a}
 \left[\partial^2_z-2w^2\partial^2_w
 -w\partial_w\partial_z+6W_0w^2\right]\vert\Psi>=0\,{,}
 \end{equation}
 \begin{equation}\label{94a}
 -i\partial_{z}\vert\Psi>=\Sigma_0\,\vert\Psi>\,{.}
 \end{equation}
 However, we have done simple factor ordering considerations in the
Wheeler-De Witt  equation (\ref{93a}). Immediately, the wave function has an
oscillatory factor, being
 \begin{equation}\label{95a}
 \vert\Psi>\sim e^{i\Sigma_0z}\vert\chi(w)>\,{.}
 \end{equation}
 The function $\vert\chi>$ satisfies the Bessel differential equation
 \begin{equation}\label{96a}
 \left[w^2\partial^2_w+i\frac{\Sigma_0}{2}\,w\,\partial_w+
 \left(\frac{\Sigma_0^2}{2}-3W_0w^2\right)\right]\chi (w)=0\,{,}
 \end{equation}
 whose solutions are linear combinations of Bessel functions $Z_{\nu}(w)$
 \begin{equation}\label{97a}
 \chi (w)=w^{1/2-i\Sigma_0/4}Z_{\nu}(\lambda w)\,{,}
 \end{equation}
 where
 \begin{equation}\label{98a}
 \nu =\pm\frac{1}{4}\, \sqrt{4-9\Sigma_0^2-i4\Sigma_0}\,{,}
 \quad \lambda =\pm 9\sqrt{\frac{W_0}{2}}\,{.}
 \end{equation}
 The oscillatory regime for this component depends on the reality
 of $\nu$ and $\lambda$. The wave function of the universe, from Noether
symmetry (\ref{86a}) is then
 \begin{equation}\label{99a}
 \Psi (z, w)\sim e^{i\Sigma_0[z-(1/4)\ln w]}\,
 w^{1/2}Z_{\nu}(\lambda w)\,{.}
 \end{equation}
 For large $w$, the Bessel functions have an exponential behavior, so that the wave function (\ref{99a}) can be
written as
 \begin{equation}\label{100a}
 \Psi\sim e^{i[\Sigma_0z - (\Sigma_0/4)\ln w\pm \lambda w]}\,{.}
 \end{equation}
 Due to the oscillatory behaviour of $\Psi$,  Hartle's criterion
is immediately recovered. By identifying the exponential factor of
(\ref{100a}) with $S_0$, we can recover the conserved momenta
$\pi_z, \pi_w$ and select classical trajectories. Going back to
the old variables, we get the cosmological solutions
 \begin{equation}\label{101a}
 a(t)=a_0e^{(\lambda/6)t}\,\exp{\left\{-\frac{z_1}{3}\,
 e^{-(2\lambda/3)t}\right\}}\,{,}
 \end{equation}
 \begin{equation}\label{102a}
p(t)=p_0e^{(\lambda/6)t}\,\exp{\{z_1\,
 e^{-(2\lambda/3)t}\} }\,{,}
 \end{equation}
where $a_0, p_0$ and $z_1$ are integration constants. It is clear
that $\lambda$ plays the role of a cosmological constant and
inflationary behavior is asymptotically recovered.

\subsubsection{The case $s=-2$}
%
The new variables adapted to the foliation for the solution
(\ref{87a}) are now
 \begin{equation}\label{103a}
 w(a, p)=ap\,{,}\qquad z(a)=a^2\,{.}
 \end{equation}
 and Lagrangian (\ref{77a}) assumes the form
 \begin{equation}\label{104a}
 \tilde{{\cal L}}(w, \dot{w}, \dot{z})=3\dot{z}\dot{w}-6kw-W_1w^3\,{,}
 \end{equation}
 The conjugate momenta are
 \begin{equation}\label{105a}
 \pi_z=\frac{\partial \tilde{{\cal L}}}{\partial\dot{z}}=3\dot{w}=\Sigma_1\,{,}
 \end{equation}
 \begin{equation}\label{106a}
 \pi_w=\frac{\partial \tilde{{\cal L}}}{\partial\dot{w}}=3\dot{z}\,{.}
 \end{equation}
The Hamiltonian is given by
 \begin{equation}\label{107a}
 {\cal H}(w, \pi_w, \pi_z)=\frac{1}{3}\,
 \pi_z\pi_w+6kw+W_1w^3\,{.}
 \end{equation}
 Going over the same steps as above, the wave function of the
universe is given by
 \begin{equation}\label{108a}
 \Psi (z, w)\sim e^{i[\Sigma_1z+9kw^2+(3W_1/4)w^4]}\,{,}
 \end{equation}
 and the classical cosmological solutions are
 \begin{equation}\label{109a}
 a(t)=\pm\sqrt{h(t)}\,{,}\qquad p(t)=\pm
 \frac{c_1+(\Sigma_1/3)\,t}{\sqrt{h(t)}}\,{,}
 \end{equation}
 where
 \begin{eqnarray}\label{h(t)}
 h(t)&=&\left(\frac{W_1\Sigma_1^3}{36}\right) t^4+
 \left(\frac{W_1w_1\Sigma_1}{6}\right)
 t^3+\left(k\Sigma_1+\frac{W_1w_1^2\Sigma_1}{2}\right)\,
 t^2+\nonumber\\&&+w_1(6k+W_1w_1^2)\, t+z_2\,{.}
 \end{eqnarray}
 $w_1$, $z_1$ and $z_2$  are integration constants. Immediately we
 see that, for large $t$
 \begin{equation}\label{111a}
 a(t)\sim t^2\,{,}\qquad p(t)\sim \frac{1}{t}\,{.}
 \end{equation}
 which is a power-law inflationary behavior.
 An extensive discussion of Noether symmetries in $f(R)$ gravity is in \cite{antoniodefelice,basilakos1}.

\subsection{ $f(T)$ cosmology}

In analogy to the  $f(R)$ gravity, a new
 sort of Extended Gravity, the so-called $f(T)$ theory,
 has been recently proposed. It is a generalized and extended version
 of the teleparallel gravity originally proposed by
 Einstein~\cite{r3,r31,r4}. 
  Teleparallelism uses as dynamical object a vierbein
field $e_i(x^\mu)$, $i = 0, 1, 2, 3$, which is an orthonormal
basis for the tangent space at each point $x^\mu$ of the manifold: $e_i . e_j=\eta_{ij}$, where $\eta_{ij}=diag(-1,1,1,1)$. Each vector $e_i$ can be described by its components $e^\mu_i$, $\mu=0,1,2,3$ in a coordinate basis; i.e. $e_i=e^\mu_i\partial_\mu$. Notice that latin indexes refer to the tangent space,
while greek indexes label coordinates on the manifold.
 The metric tensor is obtained from the dual vierbein as $g_{\mu\nu}(x)=\eta_{ij} e^i_\mu(x)e^j_\nu(x)$.
 Furthermore
 the Weitzenb\"ock connection is used, 
 whose non-null torsion is
\begin{equation}\label{torsion}
    T^\lambda_{\mu\nu}=\hat{\Gamma}^\lambda_{\nu\mu}-\hat{\Gamma}^\lambda_{\mu\nu}=e^\lambda_i(\partial_\mu e^i_\nu - \partial_\nu e^i_\mu).
\end{equation}
 rather than the
 Levi-Civita connection which is used in General Relativity.
 
 This tensor encompasses all the information about the
gravitational field. The Lagrangian is built by
the torsion (\ref{torsion}) and its dynamical equations for the vierbein
imply the Einstein equations for the metric. The
teleparallel Lagrangian is
\begin{equation}\label{lagrangian}
    T={S_\rho}^{\mu\nu}{T^\rho}_{\mu\nu},
\end{equation}
where
\begin{equation}\label{s}
    {S_\rho}^{\mu\nu}=\frac{1}{2}({K^{\mu\nu}}_\rho+\delta^\mu_\rho {T^{\theta\nu}}_\theta-\delta^\nu_\rho {T^{\theta\mu}}_\theta)
\end{equation}
and ${K^{\mu\nu}}_\rho$ is the contorsion tensor
\begin{equation}\label{contorsion}
    {K^{\mu\nu}}_\rho=-\frac{1}{2}({T^{\mu\nu}}_\rho-{T^{\nu\mu}}_\rho-{T_\rho}^{\mu\nu}),
\end{equation}
which equals the difference between Weitzenb\"{o}ck and Levi-Civita connections.
  Thus the  general action assume the following form
\begin{equation}\label{action}
    {\cal S} = \int{d^4xef(T)},
\end{equation}
where $e=det(e^i_\mu)=\sqrt{-g}$. 
If matter couples to the metric in the standard form then the variation of the action with respect to the vierbein leads to the equations \cite{Ferraro}
\begin{eqnarray}
  && e^{-1}\partial_\mu(e{S_i}^{\mu\nu})f'(T)-e_i^\lambda {T^\rho}_{\mu\lambda}{S_\rho}^{\nu\mu}f'(T) +{S_i}^{\mu\nu}\partial_\mu(T)f''(T) +\frac{1}{4}e^\nu_if(T)={e_i}^\rho {T_\rho}^\nu,\nonumber\\ \label{equations}
\end{eqnarray}
where a prime denotes differentiation with respect to $T$,
${S_i}^{\mu\nu}={e_i}^\rho {S_\rho}^{\mu\nu}$ and $T_{\mu\nu}$ is the matter energy-momentum
tensor.
 In order to derive the cosmological equations in a FRW metric, we need to infer, as above,  a point-like Lagrangian from the action(\ref{action}). As a consequence, the infinite number of degrees of freedom of the original field theory will be reduced to a finite number. Then we can write 
 
 \begin{equation}\label{metric}
    e^i_\mu = diag(1, a(t), a(t), a(t)),
\end{equation}
where $a(t)$ is the cosmological scale factor, and the action  is
\begin{eqnarray}
{\cal S}=\int {\cal L}(a,{\dot a},T, {\dot T})dt
\end{eqnarray}
considering $a$ and $T$ as canonical variables, whereas
 ${\cal Q}=\{a,T\}$ is the configuration space, and
 ${\cal TQ}=\{a,\dot{a},T,\dot{T}\}$ is the related tangent
 bundle on which $\cal L$ is defined.
As above, one can use the method of Lagrange
 mutipliers to set $T$ as a constraint of the dynamics. Selecting the suitable Lagrange mutiplier
 and integrating by parts, the Lagrangian $\cal L$ becomes
 canonical and then we have~\cite{antoniodefelice,basilakos2,nesseris,r21} 
  \begin{eqnarray}
 \label{eq12}
 {\cal S}=2\pi^2\int dt\,a^3\left[f(T)-\lambda\left(T+
 6\frac{\dot{a}^2}{a^2}\right)-\frac{\rho_{m0}}{a^3}\right],
 \end{eqnarray}
 where $\lambda$ is a Lagrange mutiplier. The variation  with
 respect to $T$, gives
 \begin{eqnarray}\label{eq13}
 \lambda=f'(T)\,.
 \end{eqnarray}
 Therefore, the action~(\ref{eq12}) can be rewritten as
 \begin{eqnarray}\label{eq14}
 {\cal S}=2\pi^2\int dt\,a^3\left[f(T)-f'(T)\left(T+
 6\frac{\dot{a}^2}{a^2}\right)-\frac{\rho_{m0}}{a^3}\right],
 \end{eqnarray}
 and then the point-like Lagrangian reads
 \begin{eqnarray}\label{eq15}
 {\cal L}(a,\dot{a},T,\dot{T})=a^3\left[f(T)-f'(T) T\right]-
 6f'(T) a\dot{a}^2-\rho_{m0}\,.
 \end{eqnarray}
Substituting Eq.~(\ref{eq15}) into the Euler-Lagrange
 equation, we obtain
 \begin{eqnarray}
 &&a^3 f''(T)\left(T+
 6\frac{\dot{a}^2}{a^2}\right)=0\,,\label{eq17}\\
 &&f(T)-f'(T )T+2f_T H^2+
 4\left[f'(T)\frac{\ddot{a}}{a}+Hf''(T)\dot{T}\right]=0\,.\label{eq18}
 \end{eqnarray}
 If $f''(T)\not=0$, from Eq.~(\ref{eq17}), it is easy to find that
 \begin{eqnarray}\label{eq19}
 T=-6\left(\frac{\dot{a}}{a}\right)^2=-6H^2\,.
 \end{eqnarray}
This
 is the Euler constraint for dynamics. Substituting
 Eq.~(\ref{eq19}) into Eq.~(\ref{eq18}) and using
 $\ddot{a}/a=H^2+\dot{H}$, we get
\begin{eqnarray} \label{eq20}
 48H^2 f''(T)\dot{H}-4f'(T)\left(3H^2+\dot{H}\right)-f(T)=0\,.
 \end{eqnarray}
 By a Legendre transformation on  Lagrangian (\ref{eq15}), we obtain the Hamiltonian constraint
\begin{eqnarray} \label{eq22}
 {\cal H}(a,\dot{a},T,\dot{T})=a^3\left[-6f'(T)
 \frac{\dot{a}^2}{a^2}-f(T)+f'(T) T+\frac{\rho_{m0}}{a^3}\right].
 \end{eqnarray}
 Considering the total energy ${\cal H}=0$ \cite{r16,antoniodefelice,r21} and using Eq.~(\ref{eq19}),
 we get
\begin{eqnarray} \label{eq23}
 12H^2 f'(T)+f(T)=\frac{\rho_{m0}}{a^3}\,,
 \end{eqnarray}
 In summary, the point-like Lagrangian(\ref{eq15})  yields all the  equations of
 motion \cite{Wei}.
 To obtain the Noether symmetries,  we substitute Eq.~(\ref{eq15}) into
 Eq.~(\ref{18}) imposing $ L_X{\cal L}=0$ and using the relations
   $\dot{\alpha}=
 (\partial\alpha/\partial a)\,\dot{a}
 +(\partial\alpha/\partial T)\,\dot{T}$, $\dot{\beta}=
 (\partial\beta/\partial a)\,\dot{a}+
 (\partial\beta/\partial T)\,\dot{T}$,
   we obtain
\begin{eqnarray} \label{eq27}
&& 3\alpha a^2\left[f(T)-f'(T) T\right]-\beta a^3 f''(T)T+\nonumber\\&&
 -6\dot{a}^2\left[\alpha f'(T)+\beta a f''(T)+
 2a f'(T)\frac{\partial\alpha}{\partial a}\right]-
 12a\dot{a}\dot{T}\frac{\partial\alpha}{\partial T}=0\,.\nonumber\\
  \end{eqnarray}
 As mentioned above, requiring the coefficients of $\dot{a}^2$,
 $\dot{T}^2$ and $\dot{a}\dot{T}$ in Eq.~(\ref{eq27}) to be
 zero, we find that
\begin{eqnarray}
 &&a\frac{\partial\alpha}{\partial T}=0\,,\label{eq28}\\
 &&\alpha f'(T)+\beta a f''(T)+
 2a f'(T)\frac{\partial\alpha}{\partial a}=0\,,\label{eq29}\\
 &&3\alpha a^2\left(f(T)-f'(T) T\right)-
 \beta a^3 f''(T)T=0\,.\label{eq30}
 \end{eqnarray}
 In particular, the constraint~(\ref{eq30}) is the {\it  Noether condition} \cite{antoniodefelice,Wei}. The corresponding
 constant of motion ({\it Noether charge}),
 reads
\begin{eqnarray}\label{eq31}
 Q_0=-12\alpha f'(T) a\dot{a}=const.
 \end{eqnarray}
A solution of Eqs.~(\ref{eq28}), (\ref{eq29}) and~(\ref{eq30})
 exists if explicit forms of $\alpha$ and $\beta$ are found. In this case, as above, a
 symmetry exists. Obviously, from Eq.~(\ref{eq28}),
 it is easy to see that $\alpha$ is independent of $T$, and
 hence it is a function of $a$ only, {\it i.e.}, $\alpha=\alpha(a)$.
 On the other hand, from Eq.~(\ref{eq30}), we have
 \begin{eqnarray}\label{eq32}
 \beta a f''(T)T=3\alpha\left(f(T)-f'(T) T\right).
  \end{eqnarray}
 Multiplying by $T$  Eq.~(\ref{eq29}), and then
 substituting Eq.~(\ref{eq32}) into it, we obtain
\begin{eqnarray} \label{eq33}
 f'(T) T\left(2a\frac{d\alpha}{da}-2\alpha\right)+3\alpha f(T)=0\,.
   \end{eqnarray}
 One can perform a separation of variables  and
 recast Eq.~(\ref{eq33}) as
\begin{eqnarray} \label{eq34}
 1-\frac{a}{\alpha}\frac{d\alpha}{da}=\frac{3f(T)}{2f'(T) T}\,.
   \end{eqnarray}
 Since its left-hand side is a function of $a$ only and its
 right-hand side is a function of $T$ only, they must be
 equal to a  constant in order to ensure that
 Eq.~(\ref{eq34}) holds. For convenience, we let this
 constant be $3/(2n)$, and then Eq.~(\ref{eq34}) can be
 separated into two ordinary differential equations, {\it i.e.},
 \begin{eqnarray}
 nf(T)=f'(T) T\,,\qquad1-\frac{a}{\alpha}\frac{d\alpha}{da}=\frac{3}{2n}\,.\label{eq36}
   \end{eqnarray}
 It is easy to find the solutions of these two ordinary
 differential equations, namely
 \begin{eqnarray}
 f(T)=\mu_0 T^n\,,\qquad\alpha(a)=\alpha_0\,a^{1-3/(2n)}\,,\label{eq38}
   \end{eqnarray}
 where $\mu_0$ and $\alpha_0$ are integral constants. Obviously,
 $f(T)$ and $\alpha(a)$ are both power-law functions. Substituting
 Eqs.~(\ref{eq36}) and~(\ref{eq38}) into Eq.~(\ref{eq32}), we
 find that
 \begin{eqnarray}\label{eq39}
 \beta(a,T)
 =-\frac{3\alpha_0}{n}\,a^{-3/(2n)}\,T\,.
    \end{eqnarray}
  Therefore a  Noether symmetry
 exists. Finally, we find out the exact solution 
 $a(t)$ for this family of $f(T)$. Substituting
 Eqs.~(\ref{eq36}), (\ref{eq38}) and~(\ref{eq19}) into
 Eq.~(\ref{eq31}), we obtain an ordinary differential
 equation for  $a(t)$, namely
   \begin{eqnarray} \label{eq40}
 a^{c_1}\,\dot{a}=c_2\,,
   \end{eqnarray}
 where
 \begin{eqnarray} \label{eq41}
 c_1=\frac{3}{2n}-1\,,~~~~~~~
  c_2=\left[\frac{Q_0}{-12\alpha_0\mu_0 n(-6)^{n-1}}
 \right]^{1/(2n-1)}\,.
   \end{eqnarray}
 It is easy to find that the general solution for  Eq.~(\ref{eq40}) is
  \begin{eqnarray}\label{eq42}
 a(t)=-(1+c_1)(c_3-c_2 t)^{1/(1+c_1)}
 =(-1)^{1+2n/3}\cdot\frac{3}{2n}\,(c_2 t-c_3)^{2n/3}\,,
   \end{eqnarray}
 where $c_3$ is another  integration  constant. Obviously, in the late
 time $|c_2 t|\gg |c_3|$ and the universe experiences a
 power-law expansion. 
 Requiring $a(t=0)=0$, it is easy to see that the integration
 constant $c_3$ is zero. Finally, we have a behavior of the form
 \begin{eqnarray} \label{eq43}
 a(t)\sim t^{2n/3}\,,
   \end{eqnarray}
  which is clearly a Friedman behavior.
 Note that the condition $n>0$ is required to ensure that the universe is expanding \cite{Wei}.
%
 
 \section{Conclusions}

 In this lecture notes, we  have discussed  the Noether Symmetry Approach   for  Minisuperspace Quantum Cosmology. The method allows  to identify conserved quantities that select  peaked behaviors in the  solutions of the Wheeler-De Witt equation.
Peaked behaviors mean correlations among variables and then the possibility to obtain classical universes according to the interpretative Hartle criterion. Specifically, such a criterion states that classical observable universes are solutions of dynamics as soon as   correlations among physical variables are identified.
Here, we search for Noether symmetries that allow to reduce dynamics coming from minisuperspaces and then find out exact solutions.

The method has been worked out for several examples of Extended Theories of Gravity, namely scalar-tensor gravity, $f(R)$-gravity and $f(T)$-gravity.
The common feature of such  dynamical systems  is that, in any case, specific Lagrange multipliers, related to symmetries, can be found out. 
Such multipliers  allow to reduce dynamics and then  exact cosmological solutions can be easily found.

Here we have worked out some very didactic examples but more physically interesting systems can be taken into account (see for example \cite{antoniodefelice, nesseris,basilakos1,basilakos2}).

\section{Acknowledgments}
We acknowledge S.D. Odintsov for useful discussions, common researches and results on the presented topics.

\end{document}